# Agentic World Analysis (AWA)

# - an alternative way to explore systems and support decision making

Yongchao Zeng [a, b,*], Alexey Voinov [c], Calum Brown [a], Tatiana Filatova [b], Mark Rounsevell [a, d, e]

a. Institute of Meteorology and Climate Research, Atmospheric Environmental Research (IMK-IFU), Karlsruhe Institute of Technology, Germany
b. Department of Multi-Actor Systems, Faculty of Technology, Policy and Management, Delft University of Technology, the Netherlands
c. University of Twente, the Netherlands
d. Institute of Geography and Geo-ecology, Karlsruhe Institute of Technology, Germany
e. School of Geosciences, University of Edinburgh, UK

* Corresponding author

Email: yongchao.zeng@kit.edu (Y. Zeng)

## Abstract

To address increasingly pressing sustainability challenges, various approaches have been developed to foresee possible futures, identify failure modes, detect vulnerabilities, and test potential mitigations. However, environmental systems are highly complex. Especially when coupled with human processes, the scale of uncertainties becomes intractable. To address this challenge, we propose a new approach – Agentic World Analysis (AWA) – combining the strengths of simulation modelling and expert elicitation. The concept of AWA is defined by three properties: 1) AWA uses an agentic AI system to mimic an expert panel that studies the world; 2) AWA projects futures iteratively through analysing scenario trees and learning from this analysis to improve decisions; 3) AWA is auditable. Based on these requirements, we implemented the World Engine by Generative Agents (WEGA) as a possible application of the AWA approach and demonstrated its functionality with a real-world case study: the Nitrogen Crisis in the Netherlands. WEGA autonomously constructed the context, identified key stakeholders and uncertainties, created expert agents, and generated future scenarios. As a result, two pathways from 2026 to 2041 were proposed, sharing a common assumption that social acceptance of nitrogen mitigation policies is low, while differing in how successful the restoration is according to the implementation of nitrogen data monitoring. The pathways are evaluated in multiple dimensions to assess their logical coherence and quality. The evaluation also actively exposes strengths and weaknesses to provide ways for testing the validity of the policies proposed. We discussed scaling up scenario analyses to enable massive pathway exploration, the trade-offs of using AWA and other approaches, and common concerns regarding AI systems.

## 1. Introduction

For centuries, fortune tellers gazed into uncertain futures through physical or figurative crystal balls. The enduring appeal of this idea is not only its mystery but also its promise of a form of anticipatory intelligence. In many ways, modern world analysis and decision-making use experts instead of fortune tellers who then work with models and algorithms, serving the same purpose as the crystal ball. In many cases, decision-makers do not know how models work or what is behind the predictions they produce; experts are still needed to ask the right questions and try to interpret the model output appropriately (Beisbart, 2021; Grimm et al., 2020; Kolkman, 2022). Indeed, shortcomings in model transparency and evaluation may mean that the experts may also not know how the models work, much like fortune tellers who did not know how the crystal ball was providing them with insights. We replace mysticism with trust and increasingly powerful computational tools for information processing. These new tools help us understand complex systems, explore uncertainty, and prepare for future outcomes. This is particularly important for understanding how human activities shape environmental change, which is a challenge at the core of sustainability science and governance (Clark and Harley, 2020; Keck et al., 2025). Human activities simultaneously impact multiple domains spanning ecosystems, economic development, power distribution in society, and long-term sustainability (Keck et al., 2025). Numerous feedbacks and interactions connect various human and natural systems in complex, nonlinear ways (Taelman et al., 2024).

Despite the decades of advancements in modelling and simulation, there are still limitations in representation, adaptability, knowledge integration, and interoperability. Model reuse and integration remain a problem. Models are always built for particular purposes, which makes it hard to use them for other systems and case studies. Learning from already available models remains a challenge because of poor documentation and long model-building life cycles. Models are often developed by different researchers from various domains, using different abstraction strategies, scales, and assumptions (Binder et al., 2013; Partelow, 2018, 2023). The increasing complicatedness in models, together with other limitations, such as code availability, transparency, and documentation clarity, all demotivate researchers from investing enough time or effort to understand others' models, leading to a fragmented modelling ecosystem (Konkol et al., 2019; Roxburgh et al., 2022; Zhu et al., 2023).

Many mechanisms, such as institutional shifts, public reactions, conflict evolution, or unexpected incidents, emerge spontaneously. Introducing new mechanisms, feedback loops, or data to better reflect the ever-changing reality requires substantial work in model modification, making it difficult to adapt to evolving research questions and real-world developments. Ideally, a model should be a living, adapting and evolving tool that is constantly updated and improved. Further, to represent the varied conditions that could arise in the future, a model itself needs to access commensurately varied structures and parameterisations even - or perhaps especially - where these are not relevant to present-day conditions. However, most models are supported by limited funding schemes that often end even before the models are brought to some level of utility. Moreover, many real-world systems are inherently difficult to formalise in models (Preiser et al., 2018; Preiser et al., 2021). A wealth of knowledge about human activities and their interactions with the environment is recorded in natural language (Blanchy et al., 2023; Smith et al., 2021), such as policy documents, social surveys, expert judgement, and stakeholder narratives. Formalising such knowledge or information inevitably omits meaningful details and results in a gap between model abstraction and actionable insights for human-environment dynamics and sustainability governance. What is also problematic is the time constraints. Developing and testing a good model may require years, and finding the right experts and learning from them may take months, perhaps also years. In the meantime, our world rapidly changes. We make decisions based on our understanding of the problem, which is already outdated.

The participatory modelling approach has been proposed as a way to address some of these concerns (Kolagani et al., 2025; Voinov et al., 2016). In this case, modelling is opened for stakeholder engagement, bringing in all sorts of participants, including experts and ordinary people who may be somehow affected. Here, the focus was shifted from building models to using the modelling process as a framework for learning, sharing, and communicating. This was supposed to help stakeholders come up with solutions that would be most acceptable and actionable, leading to better management and decision support. In some cases, this was quite successful; however, overall, the implementation of this approach remains quite limited and episodic.

Numerous expert elicitation (EE) approaches involving human participants provide much flexibility, such as Delphi (Zhao et al., 2023), Cooke's method (Krueger et al., 2012), and the Sheffield Elicitation Framework (Fovargue et al., 2019). Instead of relying on formalised knowledge (mathematical models, equations, parameters), EE leverages domain experts' knowledge, experience, and evaluation in qualitative and quantitative form to explore uncertain futures (Hemming et al., 2018). This enables EE to utilise knowledge that is difficult to formalise, account for emerging events that formal models struggle to represent, and address newly arising or poorly understood problems without laborious modelling processes. Therefore, in dealing with problems in a complex socio-environmental context, where lack of data, system evolution or critical processes involve high uncertainty, EE often demonstrates higher flexibility and adaptability. However, limitations come along with these advantages. Unlike simulations, which apply the same formal logic consistently across scenarios, expert elicitation remains methodologically heterogeneous and may lack consistent protocols, definite causal structures, or traceable processes to evaluate its quality (Krueger et al., 2012; Martin et al., 2012). Moreover, unless one does it in a participatory way, expert elicitation lacks the iterative mechanisms that characterise simulation models, which makes it difficult to propagate temporary or local changes through feedback loops or examine how such changes accumulate over time. Moreover, they may be influenced by human biases and preferences and may lack objectivity.

Recent advancements in large language models (LLMs) and generative AI agents open new opportunities that may enhance both simulation modelling and expert-based approaches. AI agents can operate via natural-language reasoning and contextualised interpretation (Sapkota et al., 2025), using the most recent and rapidly changing information. They can interact with humans through dialogues, process informalized information, and generate logically coherent narratives (Kolt, 2025). When needed, they can generate their own new models or use existing ones. This does not make them magical or flawless; these same properties mean that they can mislead and propagate errors. However, AI agents do create the possibility of a flexible approach to addressing complex problems, intelligently relying on the wealth of available data useful for conceptualising the decision logic of different actors. Recent research has demonstrated the powerful reasoning and task completion capabilities of sophisticated AI systems, such as using AI to automate scientific research (Jin, 2025), solve open mathematical problems (OpenAI, 2026), and design advanced algorithms (Novikov et al., 2025). Companies in legal, tax, accounting, and risk management are also rapidly adopting AI technologies to boost productivity and service quality (Jiang et al., 2024; Schmidt et al., 2026). There is a proliferation of various machine learning methods that generate quite accurate empirical models, which can also assist in decision-making (Farooq and Khan, 2025; Umutoni and Samadi, 2024). These examples signify that AI is rapidly expanding into domains that demand high creativity, logical rigour, and reliable reasoning, which implies that current AI technologies could be leveraged to address complex socio-environmental issues.

Agentic World Analysis (AWA), a new methodology introduced here, is established on the basis of this background. AWA is, of course, not a wizard or prophet with a crystal ball as imagined, but pursues a similar goal: meaningfully projecting futures full of uncertainty, while helping with analyses and managerial/operational decisions. It can also quickly absorb new information and adapt to it. AWA integrates the strengths of simulation modelling, expert-centred approaches, and agentic AI technologies within a unified analytical framework in response to a set of persistent challenges in socio-environmental analysis. AWA supplements traditional simulation modelling by leveraging agentic AI systems for adaptive scenario generation of future development through active reasoning. It then helps to analyse and explore these scenarios to choose the ones that properly serve the purposes of research or decision-making (Haasnoot et al., 2019). AWA autonomously constructs contextual information around a given topic, retrieves evidence, makes assumptions, produces claims, and generates reasoning chains to infer plausible futures. Its basic analytical units are expert agents with domain-specific foci. These agents can be highly modularised to assist in transdisciplinary collaboration in development and validation.

In this paper, we introduce the concept of AWA, explain its implementation principles, and demonstrate a working system with a real-world case study. We also reflect on trade-offs related to the use of AWA in comparison with some existing analytical approaches for socio-environmental research.

## 2. Methodology

### 2.1 Concept of AWA

The core idea that inspired AWA is simple: to decide about the future of a complex system and to manage it, we create AI agents that serve as experts who study a dynamically evolving problem using the tools (including formal models) of their choice to infer and analyse plausible trends and futures. AWA is built on three critical, interconnected pillars, which define AWA's conceptual core, functional architecture, and scientific requirements.

First, AWA uses an agentic AI system to replicate the expert panel that studies the world. The AI system functions as a proxy for a group of experts who explore a real-world system impacted by human activities and natural processes. For instance, when analysing how a nationwide farmer protest can impact land use management, AWA employs expert agents with domain-specific foci as consultants to infer what could happen. AWA can decide the composition of such an expert panel and which domains they should represent, or it will include particular experts if provided by the user. The farmer protests serve as a research question that requires a response from the expert agents. While the expert agents can choose to use existing models or produce additional simulations to provide their response, the modelling effort is decided by the AWA itself and is not mandatory.

**Second, AWA projects futures iteratively by generating scenario trees and exploring their various branches.** AWA can be used iteratively to infer plausible futures. For instance, we can ask the expert agents about the potential impact of a farmer protest and expect a direct answer. Meanwhile, we can also make the temporary resolution more fine-grained and controllable by asking questions iteratively. For example, we can first ask AWA what will happen within the next five years; based on the answer and possible farmer feedback, we let AWA infer what will happen within the second five years, and so on. This iterative process makes scenario projections more contextually coherent and reliable. More importantly, due to the existence of a variety of uncertainties, there exist many different scenarios and development trajectories according to how uncertainties are resolved. AWA addresses this by creating scenario trees, supporting in-depth exploration of diverse scenario branches.

**Third, AWA is auditable.** Humans can make mistakes, and so can AWA. Human experts come with their own vested interests, individual biases, priorities, and knowledge limitations. They are further influenced by political agendas, financial incentives, and power incentives. However, AWA emphasises auditability to enable thorough examination. AWA not only gives the final results but also exposes the process of how relevant results are derived. As a tool for scientific research and decision support in complex, uncertain environments, AWA pursues usefulness rather than flawlessness. To achieve this, AWA provides sufficient transparency to enable effective assessment of credibility. This requires AWA to be auditable, such as what expert agents are involved, how they are configured, what evidence is used, how logic chains are connected, and what conclusions are derived. In practice, these requirements indicate the necessity of making every step visible and traceable; implementations of AWA should actively embrace and facilitate critical scrutiny.

It is also important to note what AWA is not,: **it is NOT an approach that attempts to fill existing modelling frameworks with LLM agents.** There are increasingly more frameworks and models that have been built with a simple but ambitious target: replacing rule-based agents with LLM agents (see this survey (Mou et al., 2026)). In such systems, natural language is often reduced to proxies of rules rather than a valuable medium that conveys meaning, while system-level emergence is supposed to arise from agent dialogues instead of algorithmic operations. For instance, quite a few studies attempt to use LLM-based agents to build an artificial society, in which numerous agents mimic real-world human individuals. In such an artificial society, LLM agents make decisions and interact with one another via text, which is then supposed to lead to system-level emergence. The generated text is largely considered a by-product rather than a carrier of insights worth our careful attention. Although those approaches appear intuitive, recent critiques highlight a set of methodological dilemmas due to the incompatibility of LLMs' high expressiveness with deliberate abstraction required in modelling (Larooij and Törnberg, 2025; Zeng et al., 2026). Zeng et al. (2026) pointed out that LLMs might be more suited for tasks where the content of generated text is central rather than a by-product. Following this philosophy, AWA avoids naïve mimicry of real-world society but incorporates a few expert agents as high-level analytical units to project future scenarios.

Having these conceptual pillars introduced, we can understand the intuition behind AWA. Consider how human experts would respond to an unexpected incident, such as a nationwide farmers' protest caused by a proposed agricultural reform. Within days, economists, policy analysts, lawyers, food security experts, etc., start to retrieve information, interpret institutional contexts, compare historical events, revise hypotheses, formulate reports, and raise critical questions guiding further analysis and policy response. Ideally, their analysis is proactive, evidence-driven, and continuously updated. Their domain-specific perspectives are integrated to gain broader insights into the evolving situation. The experts may bring in some simulation tools to improve their understanding of the problem, or they may even outsource or order the development of some additional tools or data sets. As shown in Figure 1, AWA is inspired by expert analysis and uses AI agents as autonomous analytical units to explore the complex world and support decision-making.

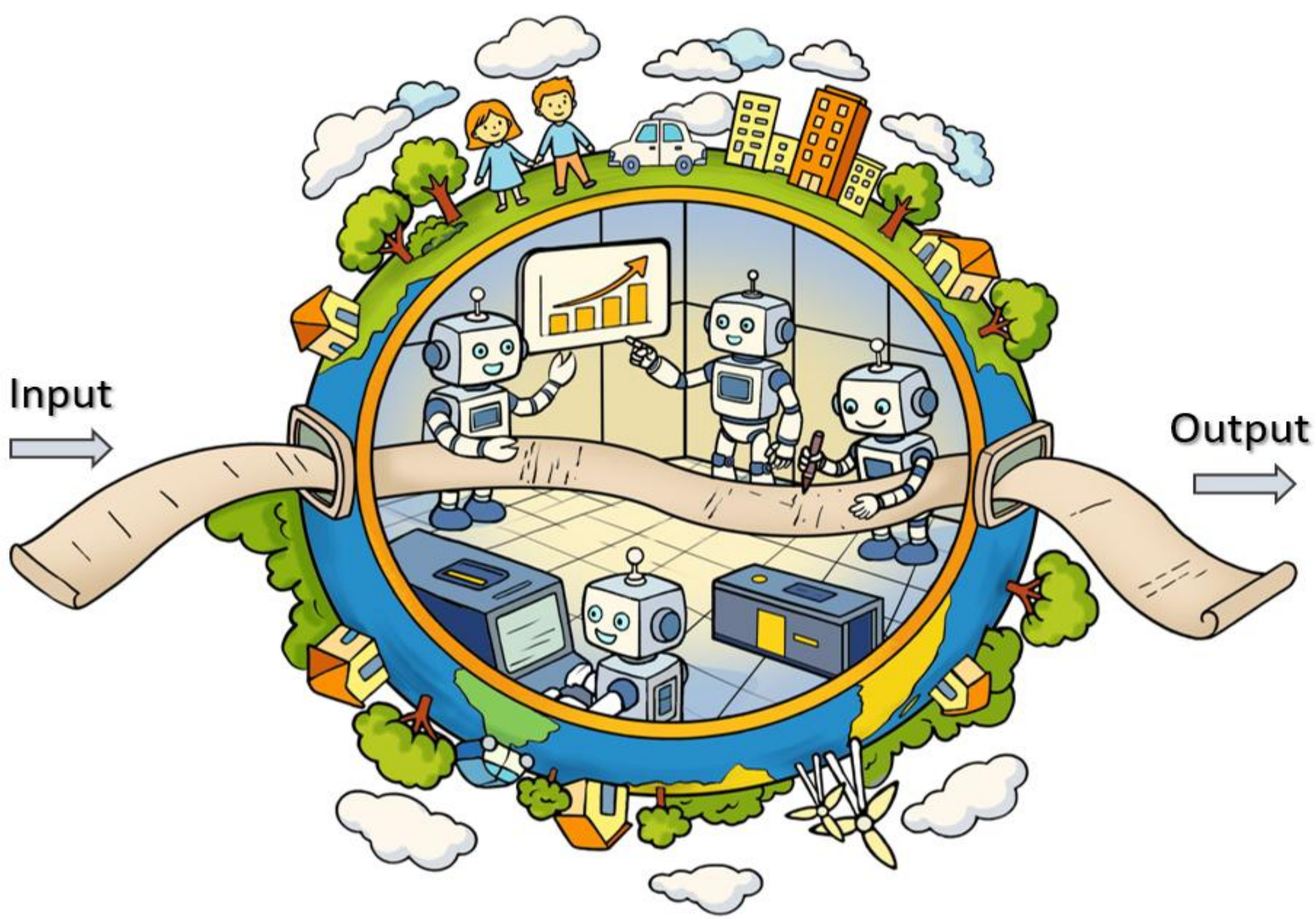

Figure 1. Metaphorical illustration of AWA. The visual metaphor highlights the central intuition while omitting many technical details: several AI agents hide inside the "Earth", busy researching answers to questions about how the world evolves. The input includes the current state of the world together with human activities or natural processes. The output contains the updated world states, which, along with newly adapted human activities or natural processes, can be fed back as input to AWA, forming an iterative loop that drives the world to evolve. Note that AWA does not simulate the real world per se, but uses AI agents as proxies of experts responding to various human concerns and problems about the world.

**2.2 World Engine by Generative Agents (WEGA) – An implementation of AWA**

We developed the World Engine by Generative Agents (WEGA) as an implementation of the AWA approach and methodology. WEGA is built mainly in Python, with access to a variety of LLM APIs, such as OpenAI and Claude models.

The overall workflow of WEGA includes four major phases: initialisation, main loop, branch scenario reloading, and evaluation (see Figure 2). Each of the phases is driven by specific AI agents, manifesting the Agent-centric design philosophy. The main features of WEGA include AI-assisted analysis configuration, automated contextualization, agent generation, branch scenario exploration, AI-assisted result analysis and evaluation. The major steps and more details about these features are as follows.

**Phase 1: Initialisation**. This phase involves four steps that prepare the initial information for the scenario projection.

> **Step 1: User input.** A user is asked to provide basic information, including the title of a case of interest, a brief description of the problem, and the time horizon that constrains the information retrieval for contextualization. Users with little knowledge about WEGA can consult the conversational agent (namely, an agentic interface), which will guide users to set up and start the analysis. Users can specify the number of iterations for the analysis. If only one iteration is required, this effectively turns the process into a question-answering machine. In addition, WEGA allows users to choose whether they want to take an active role during the scenario projection by further exploring individual scenarios among alternatives at the end of each iteration.

**Step 2: Contextualization.** A contextualization agent activates a research workflow (details in Appendix A) that collects relevant information (e.g., papers and news) online and in local storage, screens and selects the most relevant information, and extracts crucial evidence that supports building the context of analysis. The contextual information is saved as a JSON file following a predefined schema, which includes case identity, temporal context, spatial context, critical actors, core problems, etc.

**Step 3: Agent spawning.** An agent generator is responsible for creating two agent groups: one group includes key stakeholders in the case of interest, which are identified in the contextualization process; the other group comprises expert agents with domain-specific roles and foci on different analytical tasks. These expert agents can be created by WEGA according to the needs of analysis or provided as pre-defined analytical units. The generated stakeholder agents, therefore, will act as information providers, while expert agents will react to the information by analysing how stakeholder decisions impact the simulated world (more details in the "Main loop" below).

**Step 4: initial world statement.** This is the last step in Phase 1, which produces a narrative describing the initial state of the world, i.e., the background info that both the stakeholder agents and expert agents should possess.

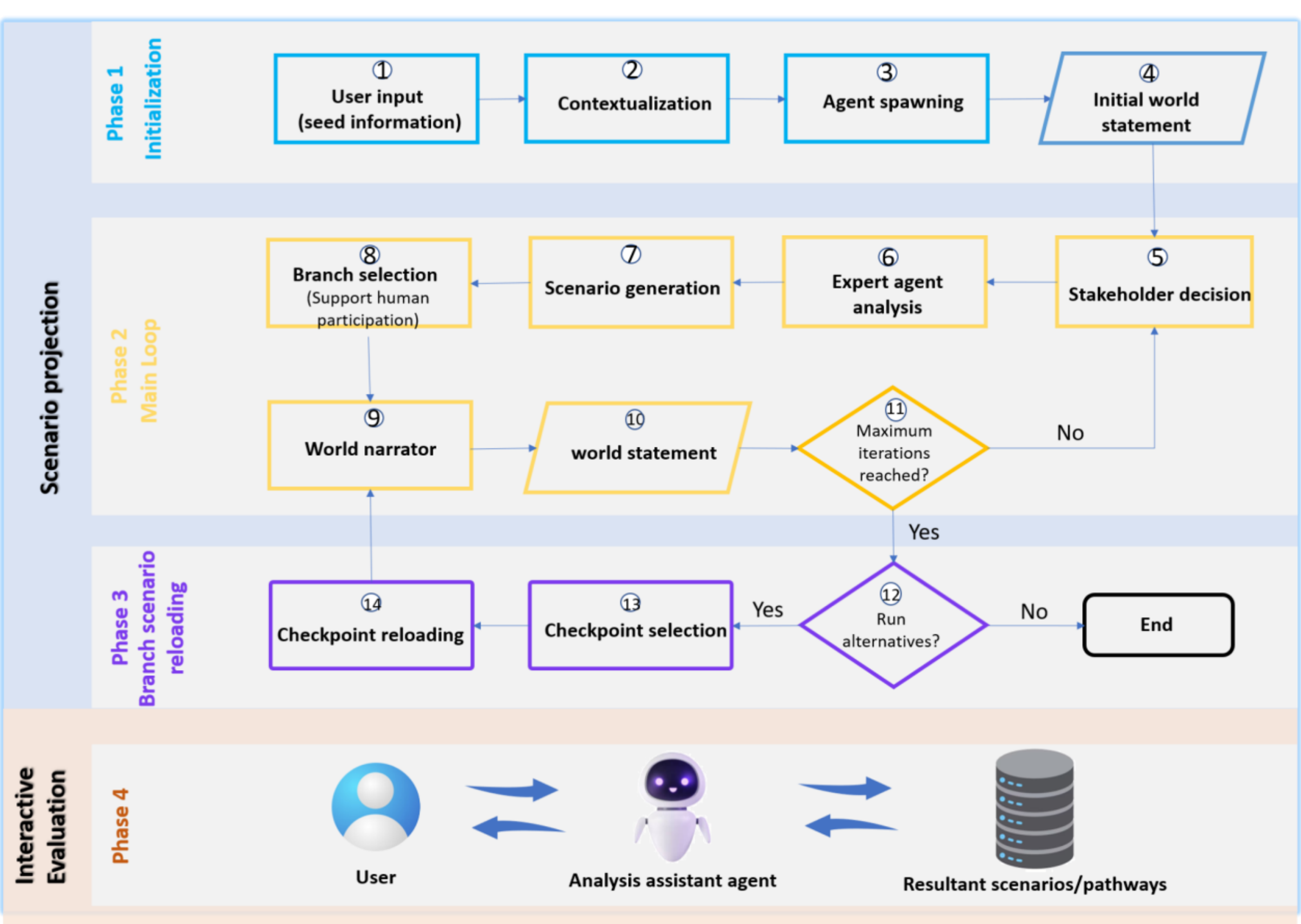

Figure 2. Diagram showing the workflow of World Engine by Generative Agents

**Phase 2: Main loop**. The iterative scenario inference processes take place in this phase, which contains Steps 5 to 11.

**Step 5: Stakeholder decision.** The stakeholder agents receive the world statement introducing the current world state and start to present their advocacies until a final decision is reached. A decision identifier driven by an LLM is responsible for extracting stakeholder decisions into a structured format instead of leaving the decision in raw text.

**Step 6: Expert agent analysis.** Both the contextual information and final decision are fed into each individual expert agent, which then starts to analyse the impact of the decision within a pre-set time window (e.g., five years). Again, the expert agents follow the same "deep research" workflow (see Appendix A) to conduct information retrieval, assumption formulation, and reasoning. Optionally, the expert agents can call existing models, similar to calling a tool, to assist with their analysis when necessary. The output of each expert agent is a file with a pre-defined structure that highlights key points, such as impact assessment, causal mechanism, and uncertainties.

**Step 7: Scenario generation.** A synthesis process is run by an agent to integrate all expert agent analyses in this iteration and generate alternative scenarios according to assumptions on how a major uncertainty is resolved.

**Step 8: Branch selection.** One of these scenarios is selected either autonomously or by a human user as the world's response to the stakeholder decision in this iteration.

**Steps 9-10: World narrator and updated world statement.** The selected scenario is broadcast as an updated world statement by the World Narrator agent. Once a scenario is selected, a checkpoint is saved to record the historical and current state of the scenario projection.

**Step 11: Stop condition check**. WEGA checks if the stop conditions are met. For instance, an easy way to end the loop is to check if a predefined number of iterations is reached. If not, repeat Steps 5 to 11; otherwise, proceed to Phase 3.

**Phase 3: branch scenario reloading.** This phase allows for the exploration of alternative scenarios generated in Phase 2.

**Step 12: Termination or branch exploration.** WEGA asks the user whether to proceed with branch scenario exploration or end the whole program. If further exploration is chosen, proceed to step 13.

**Step 13: Checkpoint selection.** The user is requested to choose from which iteration they want to rerun the scenario projection and which alternative scenario at the end of the chosen iteration should be carried forward. For instance, the user can select scenario B at the end of the second iteration because A has already been explored.

**Step 14: Checkpoint reloading.** Once a specific scenario in a specific iteration is selected, WEGA reloads the scenario projection. For instance, with scenario B in the second iteration selected, WEGA starts to infer future scenarios based on this selection, while everything before the second iteration remains the same. This will result in a new branch "storyline" with different pathways after the second iteration, which is useful for researching "what-if" questions and comparing a variety of pathways.

**Phase 4: Evaluation.** After the scenario projection is completed, all intermediate and final results are organised in a deliberately designed file structure to ease both human operators and the analytical agent to navigate through, following the state-of-the-art agent design principle – progressive disclosure. This makes the analysis easier, more precise and auditable while avoiding packing the context windows of LLMs. WEGA also provides both an analytical agent and an agent skill (i.e., "organised folders of instructions, scripts, and resources that agents can discover and load dynamically to perform better at specific tasks", according to Anthropic (2026)) that can assist users with the analysis of the generated scenarios and pathways. Users can install the agent skill on any mainstream agentic system and analyse the outcomes without the need to run AWA. The analysis can be fully interactive: users can ask open questions about the results, e.g., from the overall pathway to a specific agent's output at a specific moment. Briefly, the analytical assistant agent is guided by the following instructions:

1) Locate the run folder. The user can provide a folder of outputs for the analytical assistant agent to start with; otherwise, this agent follows a protocol to locate where the outputs are saved autonomously.

2) After locating the outputs, the agent always reads "overview.json" first. This file is generated together with other outputs. It contains the information about the scenario projection's full scope, domains, agent roles, and final outcome, enough to reason about what everything else requires.
3) Read only what the question requires. Before reading additional files, the agent is instructed to think about what the question actually needs rather than reading files speculatively.
4) Spawn sub-agents for parallel independent reads. If a question requires multiple independent files (e.g. all domain analyses for a round, or summaries across all rounds), the analytical assistant agent spawns sub-agents to read them in parallel rather than sequentially. This is an effective measure to maintain a focused context for this agent.
5) Ground every claim. Cite the specific file and field behind every statement. Explicitly distinguish the scenario states from what is inferred by comparing across files.

Thus, from analysis setup to result evaluation, all steps and outputs are recorded in a structured manner. This can facilitate both human examination and AI-assisted inspection, which reflects AWA's emphasis on auditability.

## 3. Case study

To demonstrate the methodology of AWA and the WEGA system, we use the Nitrogen Crisis in the Netherlands as a case study. This case was chosen because 1) there exists rich and specific background information recorded in academic papers, grey literature, and news; 2) it is one of the most severe ongoing environmental problems faced by the Netherlands and the EU; 3) it involves diverse stakeholders, multiple governmental bodies, and entangled environmental, social, and economic challenges.

First, WEGA asks for the setup information given in Table 1. We chose the analysis to cover the period 2026–2041 across three five-year deliberative iterations. The contextualization agent conducts research and builds the context of this use case. A brief description of the "Problem Context" derived through the contextualization is shown in Box 1. A detailed technical report on the contextual information generated by WEGA is provided in Zeng (2026a), with 15 highly relevant peer-reviewed papers cited. Based on the collected evidence, WEGA identified five stakeholders as critical in this case: a) a Dutch livestock farmer union representative, b) a European Commission official, c) an environmental NGO scientist, d) a right-wing media editor, and e) the Dutch Minister of Agriculture, Nature and Food Quality. For simplicity, corresponding stakeholder agents are then generated autonomously by the system. The case study schedules the conversation of these stakeholder agents as a turn-based process: each stakeholder agent presents its views and positions one after another; a final decision-maker must speak last and make policy adaptations based on the current world state and other stakeholders' advocacies. Here, the Dutch Minister of Agriculture, Nature, and Food Quality acts as the final policy decision-maker in each iteration (i.e., every five years).

In this case study, we pre-define three domain-expert agents providing analytical grounding regarding the social, environmental, and economic aspects for the synthesis agent to generate plausible futures under the impact of the policy decisions. We limited the number of expert agents here just to avoid massive token consumption while still preserving heterogeneity in expert agents. The results of the analysis contain a main branch and a fork branch (see Figure 3). Although WEGA can generate numerous pathways efficiently, we deliberately limit the generated branches because the demonstration purpose requires clarity over comprehensiveness. All relevant details about the analysis settings and outputs are accessible at Zeng (2026a). A human-friendly, interactive HTML file regarding the scenario analysis is provided at https://yczen.github.io/AWA_output/. Here, we briefly present how the nitrogen crisis evolves along the main and fork branches, as well as the evaluation of results provided by the analytical assistant agent.

Table 1. Initial settings of the Analysis of the case study

| Initial settings | Specifications | Remarks |
|---|---|---|
| Case title | Nitrogen Crisis in the Netherlands | The title of the case of interest. Also important for managing scenario projection and branch scenarios. |
| Case description | Agricultural nitrogen emissions exceeding EU limits, threatening biodiversity and farming livelihoods | A description of the case of interest, which can be brief or detailed, guides the contextualization process. |
| Geographical area | The Netherlands | This influences how contextualization and expert agents search for evidence related to specific areas. |
| Context period | 2010–2026 | This defines the time range for searching for contextual information. |
| Start year | 2026 | The initial year of the scenario projection. |
| Time resolution | 5 | Indicating how many years in reality one iteration in the iterative scenario projection represents. |
| Number of rounds | 3 | The number of iterations after which the main loop stops. |
| Policy focus | None | Specifying the policy focus of contextualization. |
| Interactive scenario selection | y / N | Choose whether it allows a human user to select between branch scenarios after each iteration during the scenario projection. If "N" is chosen, scenarios will be randomly selected. |

Box 1. A brief description of the "Problem Context" regarding the Nitrogen Crisis in the Netherlands. This is a part of the output produced by the contextualization agent.

> Core Problem: Persistent exceedance of EU nitrogen limits in the Netherlands due to intensive livestock and fertilizer use, creating a dual crisis: (1) Ecological: Degradation of Natura 2000 sites, groundwater contamination, and transgression of planetary boundaries for biogeochemical flows. (2) Socio-Political: Intense conflict over livelihoods, characterized by organized science denial, protests, and political polarization. The problem is structurally embedded in a high-productivity agricultural model that is incompatible with current environmental safety thresholds.

### 3.1 The main branch

In Figure 3, the main branch is illustrated with the blue circles and arrows. Starting from 2026, the contextualization process built an initial context based on the settings shown in Table 1, following the research workflow (Appendix A). The context captures a "dual crisis" – ecological and socio-political conflicts – in the Netherlands due to the incompatibility between a high-productivity model and environmental safety thresholds. To address this issue, the stakeholder agents generate a set of policies (Decision Set 1 in Table B1), such as transfer payments to support nitrogen transition, implementation of zone-specific reduction targets, mandatory publication of raw deposition modelling code, datasets, and critical load assumptions to ensure methodological transparency. The five-year impact of these policies was analysed by the expert agents. Two plausible scenarios were generated according to whether the social acceptance of these policies is low or relatively high. The system chose to proceed with the low social acceptance assumption according to the pre-set random selection mechanism. The selected scenario is characterised by a deteriorating situation, e.g., "widespread non-compliance", monitoring equipment being sabotaged, and falsified reporting. The influence of "right-wing media and populist politicians" intensifies the conflicts between farmers and "elites".

Entering 2031, the inherited world state was an entrenched institutional gridlock: systematic monitoring, sabotage, legislative deadlock, and pressing EU infringement procedures. For that, stakeholders came up with a set of "conditional de-escalation" measures (see Decision set 2 in Table B1). Among them, the most crucial policy is the "Structured Nitrogen Transition Accord (SNTA)". SNTA replaced "punitive paralysis with legally compliant implementation: if the monitoring system can be restored within 30 days and pass an independent audit, the "Dutch-EU Agricultural Transition Resilience Fund" will offer financial support with a 90-day pause on "herd-reduction orders". The system assessed two possible scenarios depending on whether monitoring integrity can be restored. The system chose the "Successful Restoration" branch. Under this scenario, farmers chose to obtain the Resilience Fund and stopped sabotage behaviour; within local cooperatives, peer pressure helped to address isolated incidents. Real-time emission dashboards accessible to individual farms drove the debate from "ideological denial to technical problem-solving".

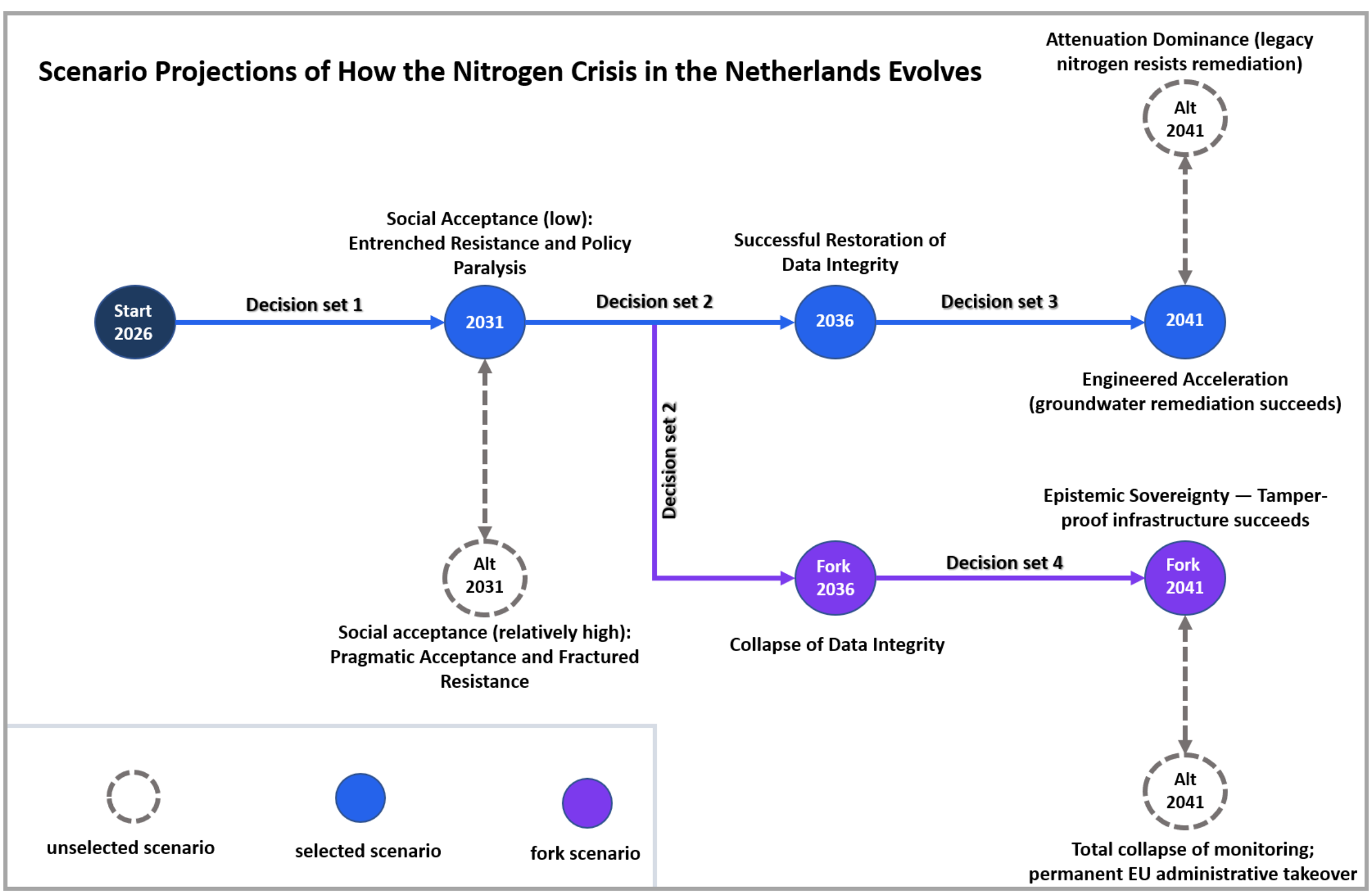

Figure 3. Pathways of how the Nitrogen Crisis in the Netherlands evolves using WEGA. An interactive version is available at https://yczen.github.io/AWA_output/

By 2036, the world demonstrated a stabilised but still fragile equilibrium state. The stakeholders proposed a series of commitments under "the SNTA Addendum" (Decision set 3 in Table B1). These included "Trajectory-Conditional Structural Guarantee", suspending mandatory herd reductions provided a verified biannual deposition trajectory toward Natura 2000 critical loads was maintained, an additional Resilience Fund tranche with 30% ringfenced for legacy groundwater remediation, and the formal open-sourcing of Joint Nitrogen Attribution & Mitigation Review (JNAMR, see Decision set 2 in Table B1) deposition models. The expert agents assessed two possible future scenarios depending on a critical uncertainty: "Efficacy of Legacy Groundwater Remediation". The selected scenario labelled "Engineered Acceleration" assumes that "legacy remediation technologies successfully intercept and process subsurface nitrogen loads". In contrast, the other scenario assumes that legacy remediation technologies failed to make a significant change in nitrogen removal, which eventually led to the triggering of the mandatory adjustment protocol.

### 3.2 The fork branch

The three iterations above comprise a complete pathway, which we can call "the main storyline". After the main storyline finished, the WEGA system asked the user if they wanted to choose previously unselected branches. To demonstrate this feature, we selected the alternative branch at the end of the second iteration (year 2031), which was labelled "Collapse of Data Integrity". In addition to the newly selected scenario, the system also loaded the complete history before the scenario was selected. This scenario described a pessimistic situation relative to the main storyline. SNTA's 30-day monitoring restoration failed within weeks. An independent nitrogen audit was framed by some media as "digital enclosure", stimulating widespread intimidation of auditors and jamming of the sensor network, which paralysed JNAMR and the Dutch-EU Agricultural Transition Resilience Fund. The EU Commission lost patience with the Dutch deadlock and initiated billions of euros in fines while suspending Common Agricultural Policy payments to the Netherlands. The Dutch government's inability to enforce domestic compliance and the external punitive measures imposed by the EU Commission further radicalised the farming community in the Netherlands.

Entering 2036, to resolve the law enforcement issue, the Dutch Minister of Agriculture, Nature and Food Quality issued a policy portfolio (Decision set 4 in Table B1), including a new framework to "reassert domestic executive control and establish compliance pathways", conditional liquidity facility for the agricultural sector, convening of an independent audit commission, sensor network integration, and required submission of "binding farm-level Voluntary Reduction & Compliance Agreements (VRCAs)". A major uncertainty estimated by the expert agents is whether this policy portfolio can restore "Epistemic Sovereignty & Data Integrity". The optimistic future scenario indicates that "a trusted, tamper-proof national monitoring infrastructure" was established, "breaking the cycle of science denial and enabling verified compliance". On the contrary, the pessimistic future scenario showed that the policy portfolio failed to restore farmer trust and data integrity; sabotage on the sensor network continued.

### 3.3 Agentic evaluation of the scenario projection

To facilitate the analysis of the results, which are mainly narratives in this case, we developed an agent skill named "analyze-awa" (Zeng, 2026b) that allows for analysing and evaluating the results interactively. The skill is employed with Claude Sonnet 4.6, which is different from the LLM (qwen-3.5-max) used to drive the iterative scenario generation. The full evaluation dialogue is available at Zeng (2026a). Here, we briefly introduce how the interactive evaluation worked and what we found.

After installing and activating the skill, we asked Claude to plan for the evaluation. It proposed seven dimensions, such as narrative coherence across rounds, branch generation quality and domain analysis quality (Table 2). Then, it showed the strategy for navigating the folders and files that contain the info generated throughout the scenario projection. For instance, the evaluation agent stated: "I'll read context.json and the three round summaries in parallel first, then selectively pull domain and stakeholder files based on what I find." Following this, it proposed an output structure, which includes four aspects: what is observed, citation of the relevant file and field, a three-level verdict (strong/adequate/weak) per dimension, and a brief overall assessment at the end. The evaluation results are summarised in Table 2.

Table 2. Evaluation results generated by Claude using the analyze-awa skill

| **Dimension** | **Verdict** | **Strengths and Limitations** |
|---|---|---|
| Narrative Coherence Across Rounds | Strong | - Strength: "The causal chain across all three parent rounds holds together convincingly".<br>- Limitation: coherence gap worth noting: despite the legislative gridlock, all policies were enacted solely by the Dutch Minister of Agriculture |
| Branch Generation Quality | Adequate, with structural constraint | - Strength: "Each selected axis is well-motivated and genuinely consequential."<br>- Limitations: Structural constraints; multiple uncertainty axes are detected in each iteration, but only two branches are available for selection. |
| Domain Analysis Quality | Strong | - Strength: "The three domain agents are genuinely differentiated and independently grounded."<br>- Limitation: Not mentioned. |
| Stakeholder Representation Quality | Strong | - Strength: farmer union representative, right-wing media, and environmental NGO scientist show "sharp differentiation both in epistemic stance and rhetorical mode".<br>- Limitation: stakeholders presented their advocacies in turn, rather than exchanging ideas in a live dialogue. |
| Evidence-Groundedness | Strong | - Strength: "context.json is the strongest single file in the simulation." Citations are explicit, including precise quantitative values; confidence levels were also estimated.<br>- Limitation: Not mentioned. |
| Fork Design Quality | Adequate, with an architectural note | - Strength: well-motivated; picked up the most consequential branch that the parent run escaped.<br>- Limitation: "only two rounds are counterfactually new" |
| Epistemic Calibration | Adequate, with a convergence concern | - Strength: "The uncertainty and unresolved_conflicts fields are consistently substantive, not boilerplate."<br>- Limitation: "both runs end with stabilisation by 2041"; unselected branches demonstrated failure plausibly but never carried forward; optimism bias. |

The evaluation results also include a summary of practical insights (see Table 3) and critical guidance on the use of the results of the scenario projection (see Box 2).

Table 3. Insights generated by Claude using the analyze-awa skill

| Insight | Portability |
|---|---|
| Monitoring integrity must precede other instruments. | High – applies to any compliance-dependent transition policy. |
| Transparency backfires in deep-trust-deficit contexts unless co-designed | High – documented beyond the Netherlands case |

| Insight | Portability |
|---|---|
| Groundwater/legacy load lags require explicit administrative decoupling | Medium – specific to soil/water pollution contexts |
| Science denial is endogenous; validator identity matters more than evidence quality. | High – applies broadly to contested-science governance. |
| Technical coercion and cooperative trust are viable alternative recovery paths with different residual costs. | Medium – depends on how far trust has already collapsed. |
| EU/external pressure threshold requires domestic trajectory credibility to function as a lever rather than an accelerant. | Medium – specific to multi-level governance contexts |

Box 2. Critical guidance on the use of the insights in Table 3, generated by Claude using the analyze-awa skill

**On how these insights should and should NOT be used**

AWA produces scenario inference, not probability estimates. The insights are structurally coherent patterns supported by the evidence base — not predictions about what will happen. The practical value is in identifying mechanisms, preconditions, and failure modes, not in reading off outcomes.

Three cautions from the evaluation findings:

**Do not read the stabilization of both runs as a prediction of convergence.** Both runs happen to end in managed stability by 2041 through editorial choices in branch selection. The unselected branches across all three rounds include total governance collapse, EU direct rule, and irreversible biodiversity loss. Those are also coherent futures; they simply weren't the paths carried forward.

**Do not over-index on the specific policy designs the simulation generates.** Mechanisms like the "Joint Nitrogen Attribution & Mitigation Review" or "Voluntary Reduction & Compliance Agreements" are illustrative devices, not blueprints. The simulation's value is in the structural conditions that make or break those mechanisms — co-design, verified data, transparent conditionality — not in the specific institutional forms.

**Do not conflate high narrative coherence with high predictive reliability.** The simulation's scenarios are internally consistent and evidence-grounded. That is the right validity criterion for this kind of tool. It does not mean the specific tipping points, timelines, or recovery speeds it projects are calibrated estimates. The groundwater remediation "40% better than RIVM models" figure in the parent R3 is a scenario assumption, not a quantitative prediction.

# 4. Discussion

## 4.1 Reflection on AWA's properties

AWA represents a new approach to problem-solving, management and decision-making. The WEGA is a concrete implementation of AWA. The case study demonstrates some key properties of AWA.

First, AWA does not rely on predefined mechanistic details. As witnessed in the case study, the WEGA system only required a small amount of seed information to be activated. As for the scenario projection of the Nitrogen Crisis in the Netherlands, we did not pre-set how the crisis should evolve or specify what causal paths should be influential. Instead, the system autonomously identified several critical mechanisms, including statements such as "science denial narratives harden despite transparency measures" and "political polarisation results in legislative gridlock". This feature enables AWA to avoid the structural rigidity commonly associated with many existing approaches; the system's internal workflow does not need to be fundamentally adapted for new domains or different problems.

Second, AWA does not rely on a strict formalisation. Due to the ability of AI agents to process both qualitative and quantitative information, AWA remains highly flexible as an analytical framework. This is particularly important for studying real-world socio-environmental problems because massive human knowledge, institutional dynamics, political narratives, and cultural factors are difficult to encode in mathematical or algorithmic forms (Rastogi et al., 2026; Yoshida et al., 2024). In the case study above, the retrieved information comes from peer-reviewed academic papers and online news. Such contextual details and evidence sources are often omitted in formal models due to the necessity of simplification and abstraction. In AWA, details that are difficult to formalise can be preserved and remain analytically influential in scenario projections. Such a detail-affinity property aligns naturally with generative AI's capability (Zeng et al., 2026). More importantly, this property enables AWA to better preserve the complexity of real-world systems while remaining closer to the communication medium of policymakers and decision-makers, who primarily use natural language to reason, argue, debate, explain, and persuade.

Third, AWA's agent-centric design philosophy makes it very suitable for collaborative development. The whole workflow is executed by a group of AI agents who natively support communication in natural language. This brings a prominent advantage to system extensibility: a single agent can be independently developed, modified, and replaced without requiring developers to deeply understand or re-engineer the rest of the system. This idea resonates with Reusable Building Blocks (RBBs), a modelling practice increasingly adopted in the agent-based modelling community (Berger et al., 2024; Filatova et al., 2025). In the case study, there are three expert agents activated. Although these expert agents shared the same deep research workflow, they were guided by different analytical foci and domain perspectives. Other researchers or developers can add, remove, or customise agents for varied analytical depth and breadth according to their research purposes and analytical requirements.

Moreover, the evaluation agent integrated within WEGA enables a self-reflection process rarely seen in existing analytical frameworks. This is important because it facilitates extracting insights and exposing vulnerability actively rather than waiting passively for post-analysis scrutiny. It can help researchers form a fair view of the analytical results and encourage reasonable use. As seen in Tables 2 and 3, as well as Box 2, the evaluation agent assessed the analytical process via multiple dimensions, highlighted key insights, and listed caveats regarding how the results should be used. The statement – "The practical value is in identifying mechanisms, preconditions, and failure modes, not in reading off outcomes." – is a meaningful statement that highlights the bounded validity scope of the analysis. It is noteworthy that the limitations identified by the evaluation agent are accurate. The fixed stakeholder roles, their turn-based conversations, and a limited number of scenarios are weaknesses because they serve the demonstration purpose, which prioritises simplicity and clarity. A formal analysis should involve varied stakeholder roles, different forms of stakeholder communication, and a large number of scenarios. This can be achieved efficiently through parallelised analytical processes.

We envision a workflow for a formal analysis using AWA, which is briefly introduced here:

1. **Problem definition**. This involves specifying the target problem for analysis and defining the time and geographical scope of the problem.
2. **Analysis setup**: Initialise the analysis by setting "seed information", similar to Table 1.
3. **Large-scale analysis.** Run the analysis, which will generate numerous scenarios and pathways in parallel based on different combinations of assumptions regarding how uncertainties are resolved.
4. **Scenario/pathway clustering**. Run text mining and analysis to tease out different families of scenarios and pathways.
5. **Scenario/pathway sampling.** Sample a few scenarios or pathways from each cluster based on research requirements. For example, robust decision-making often requires identifying the worst cases, which means the "worst scenarios" should be given higher weights during sampling.
6. **AI-assisted evaluation/reflection.** Use AI to conduct an in-depth analysis of the samples to detect, e.g., logical coherence, assumption sensibleness, and evidence soundness while highlighting both useful insights and weak points from the analysis.
7. **Expert audit.** Human experts review the sampled scenarios/paths, evaluate their validity, and write guidance on the responsible use of the results.
8. **Application.** Use the results to inform real-world decision-making.

### 4.2 What about validation?

Due to the high non-linearity, adaptability, uncertainty and partially unpredictable nature of complex socio-environmental systems, validation is a long-standing challenge in this domain (Brown et al., 2023; Petty, 2018). However, it is not quite obvious what validation implies in future-oriented analyses. We normally validate a model, but scenario runs are something we cannot really validate until we are in the future and see what really happened. Paradoxically, even then, validation is unclear, since the real system can be influenced by the scenario projections and will develop differently than if the projections were not there. Despite the paradox, it is understandable that people still tend to anticipate a form of validation that can demonstrate credibility.

Fortunately, AWA does not make validation more difficult than many existing approaches do. For example, modelling often involves unspoken or undocumented assumptions in, e.g., abstraction, parameterisation, or software implementations, making it difficult for external researchers to fully inspect or understand (Bistline et al., 2021; Hermann and Fehr, 2024). In practice, models become highly specialised and inaccessible outside the development teams. Technical barriers make these limitations even harder to expose or assess by domain experts who are not necessarily modellers. This, in effect, limits transparency, interoperability, and broader community engagement even if the model code and dataset are open-sourced.

In contrast, AWA is potentially more validation-friendly. The emphasis on accessibility and communicability enables full inspection of internal analytical coherence, from evidence grounding, assumption plausibility, logic chains, to conclusions. Following the auditability-first principle, AWA preserves all intermediate and final outputs that reflect analytical processes. These outputs are largely not in an abstract form; instead, they are human-accessible/-communicable narratives. This accessibility and communicability are crucial for the analytical processes to be properly exposed and discussed by the scientific community, which can further help to extract insights, identify flaws, probe the validity scope, and facilitate the establishment of consensus or highlight controversies.

### 4.3 Composability, distributed validation and development

A key feature of AWA is composability. For instance, the expert agents can be designed as modular components, developed, improved, or replaced by different researchers or organisations. This composability endows AWA with a unique strength: rather than concentrating the responsibility of model development and validation within a single group, AWA supports collaborative model development and distributed validation. Researchers can contribute different expert agents that represent different sets of assumptions, workflows, and theoretical perspectives. Expert agents that aim at similar tasks can compete based on performance, accuracy, ease of use, and robustness. These dimensions can also be evaluated by broad community members. Such a modelling ecosystem may have the potential to foster a more open, iterative form of model development, similar to collaborative software development or open scientific infrastructure.

This composability can also help to address the persistent interoperability and reusability issues in the modelling community. Conventional socio-environmental models are often hard to integrate because they have incompatible structures, mechanisms, assumptions, development environments, or domain-specific abstractions. Hence, researchers are frequently motivated to develop new models rather than invest effort in understanding, reusing, or enhancing existing ones. Communications across modelling communities are still limited, even though there are already considerable overlapping objectives.

Because AWA structures analysis via modular AI agents and interpretable outputs, it can substantially lower the barriers for collaboration and knowledge exchange. Researchers can contribute new analytical units without rebuilding the whole framework. In the meantime, experts who are not model developers can engage more directly in the inspection of assumptions, evidence, and reasoning trajectories. This accessibility may improve interdisciplinary communication and facilitate the integration of knowledge, methods, and perspectives within socio-environmental system analysis.

### 4.4 Common concerns about LLM-based systems and pragmatic choice of tools

Although AWA does not rely on any concrete form of AI, the current implementation – WEGA – leverages the power of LLMs. As LLMs have been rapidly adopted in a variety of domains, people have started to form different "beliefs" and "understandings" regarding LLM-driven systems or workflows. It is useful to discuss some of the common concerns.

**"LLM-based systems relying on "black-box" neural architectures: their internal representations are difficult to interpret".** This is partially true. However, the relationship between internal computational opacity and external practical interpretability is often misunderstood. Human cognition is also established via biological neural systems, which are not clearly interpretable at the mechanistic level. Nevertheless, humans can still communicate with one another using explicit language and behaviour, which can be questioned, challenged, and falsified. Analogously, although how LLMs work internally is not necessarily fully clear, AWA's analysis is externalised as interpretable reasoning, explicit assumptions, evidence retrieval, and narratives.

It is worth noting that research reveals human cognitive processes are influenced or even structured by words that "go together" – a parallel to the way LLMs work (Carruthers, 2002). In addition, at some level, human language also reflects an approach involving linguistic pattern matching, similar to LLMs (Muñoz-Ortiz et al., 2024). In reality, AWA is no more "black-box" than human reasoning because the reasoning in AWA can be well documented and reported. In practice, users and researchers do not need to know which neurons are activated or how they are connected to understand whether an LLM-generated argument is logical or evidence-grounded. Operationally, the key is whether the generated outputs can be accessed, discussed, falsified, and improved.

**"LLM outputs are not reproducible".** This is an intuition-induced misunderstanding: LLMs are stochastic word-producing machines (Sivakumar et al., 2025), so it is almost impossible to reproduce the same text. In essence, LLM outputs follow a probability distribution formed through training (Renze, 2024; Vaswani et al., 2017). This means their outputs are statistically reproducible. Indeed, LLM outputs are much more reproducible than human-involved analytical processes, such as expert group decision-making. Because the universe prohibits time travel, we cannot reset experts' brains or the world they reside in, but this does not negate the usefulness of expert-centred methods in specific situations. The key relevant question is not about reproducibility per se, but about whether we really need to reproduce the results.

WEGA generates structured narratives with evidence, assumptions, and claims traceable. People can evaluate AWA outputs by reading generated narratives, similar to reviewing an academic paper or technical report. We do not assess the value of a paper by questioning whether the author can reproduce the same text. Instead, our concern is whether the message delivered through the text is meaningful, the evidence is strong, the assumptions are plausible, or the reasoning is logical. These aspects reflect a set of shared requirements in future-oriented approaches or tasks.

When talking about LLMs, reproducibility is often brought up together with validation. It is worth noting that validating AWA outputs by reproducing the "history" is very challenging, if not impossible, e.g., reproducing what happened about the nitrogen crisis in the Netherlands between 2020 and 2026 based on the knowledge by 2019. Because there is no guarantee that relevant knowledge after 2019 has not leaked into LLM training data. A potentially sensible way to conduct such "out-of-distribution" validation is to strictly control the scope of the training data, which is very expensive and difficult to operationalise. However, whether such reproducibility is important depends on the specific analysis purposes. In many situations, understanding failures and identifying vulnerabilities are more meaningful than pursuing reproducibility when dealing with complex systems. Note, however, that this is not any different from what we currently have with our existing expert-based decision-making.

**"LLMs hallucinate and produce biased content".** These are some inherent limitations of LLMs, which might be impossible to fully eliminate, despite rapid LLM improvement and increasingly more mitigation measures (Alansari and Luqman, 2026; Joshi, 2025). Rather than taking a defensive position, it is more constructive to acknowledge these limitations and think pragmatically. After all, building models and selecting tools are about making trade-offs. Indeed, many long-existing approaches also have comparable limitations, yet with different names. For instance, expert-centred approaches inevitably involve human-related imperfections, including overconfidence, tacit knowledge, political biases, and fatigue-induced unreliability (Dror, 2020; Sanchez and Dunning, 2023). On top of that, we are dealing with humans, who have their emotions, "likes" and "dislikes", mood swings, cultural backgrounds and drivers. On the one hand, all these can be very important to consider for the decision-making process. But on the other hand, they may be no different from the hallucinations that we find in LLMs. Modelling can also inherit such limitations from modellers, but under the disguise of formalism (Saltelli et al., 2025). We just need to learn to deal with them.

AWA tackles LLM hallucination and biases strategically by 1) imposing forced reasoning, evidence grounding, and assumption explicitness throughout the analysis process, 2) operationalising auditability to actively expose problems, and 3) inviting distributed validation and improvement. Following the discussion, we compared several widely used methods in socio-environmental studies with AWA, as shown in Table 4. Neither these methods nor these dimensions are exhaustive here, and the comparison serves as a reminder of the trade-offs in method selection. It is noteworthy that AWA is envisioned as an inclusive, rather than exclusive, framework, which could be used together with other approaches. For instance, wrapping ABMs or IAMs as tools callable by expert agents in AWA is viable and can provide model-based information in addition to qualitative reasoning. Expert panels should also be integrated into AWA's scenario selection and output audit.

Table 4. Trade-off between different methods for socio-environmental studies

| Dimension | AWA (as described) | IAMs | ABMs | Expert panels |
|---|---|---|---|---|
| Quantitative output | Possible via model calls or agentic tool use | Native | Native | Possible via using models and data |
| Qualitative output | Supported | Possible if interpreted | Possible if interpreted | Supported |
| Political/social dynamics | Endogenous | Exogenous | Almost exogenous | Mainly static; authentic but biased |
| Cross-domain synthesis | Strong | Moderate to strong | Moderate | Moderate to strong |
| Validation pathway | Distributed, modular | Centralised, aggregate | Mainly centralised | None |
| Bias manageability | Replaceable components | Locked in structure | Locked in rules | Largely invisible |
| Model development | Open, distributed | Proprietary/closed | Mixed | Not applicable |
| Configurability | High | Low | Moderate | Low |
| Reproducibility | Statistically reproducible | High | High | Impossible |

## Conclusions and open questions

Agentic World Analysis offers an alternative approach to decision-making and environmental management. We tap the power of AI agents to explore and compare the possible futures to produce optimal scenarios for future development. In the meantime, the research is still in the early phase, and some questions should be tested and understood before we can fully rely on such tools. There are also a variety of new opportunities that we can explore with this system. For example, we can find out how the results are influenced by agent skills or context.

Probably a key challenge that remains and requires close attention is the relationship between AWA and the human users of the system. How can human experts and stakeholders participate in the decision-making process? If AI agents are becoming superior to what human experts can contribute, will there still be a need for human experts to be part of the game? But if humans are phased out of the process, who sets the goals, and who defines the purposes of the studies? Who decides which scenarios and decision trees should be chosen? How do we know that humans will benefit from the decisions proposed?

## Appendix A

As illustrated in Figure A1, the research process is divided into three phases, i.e., Search, Read, and Write, to streamline the agentic analysis, which is detailed as follows.

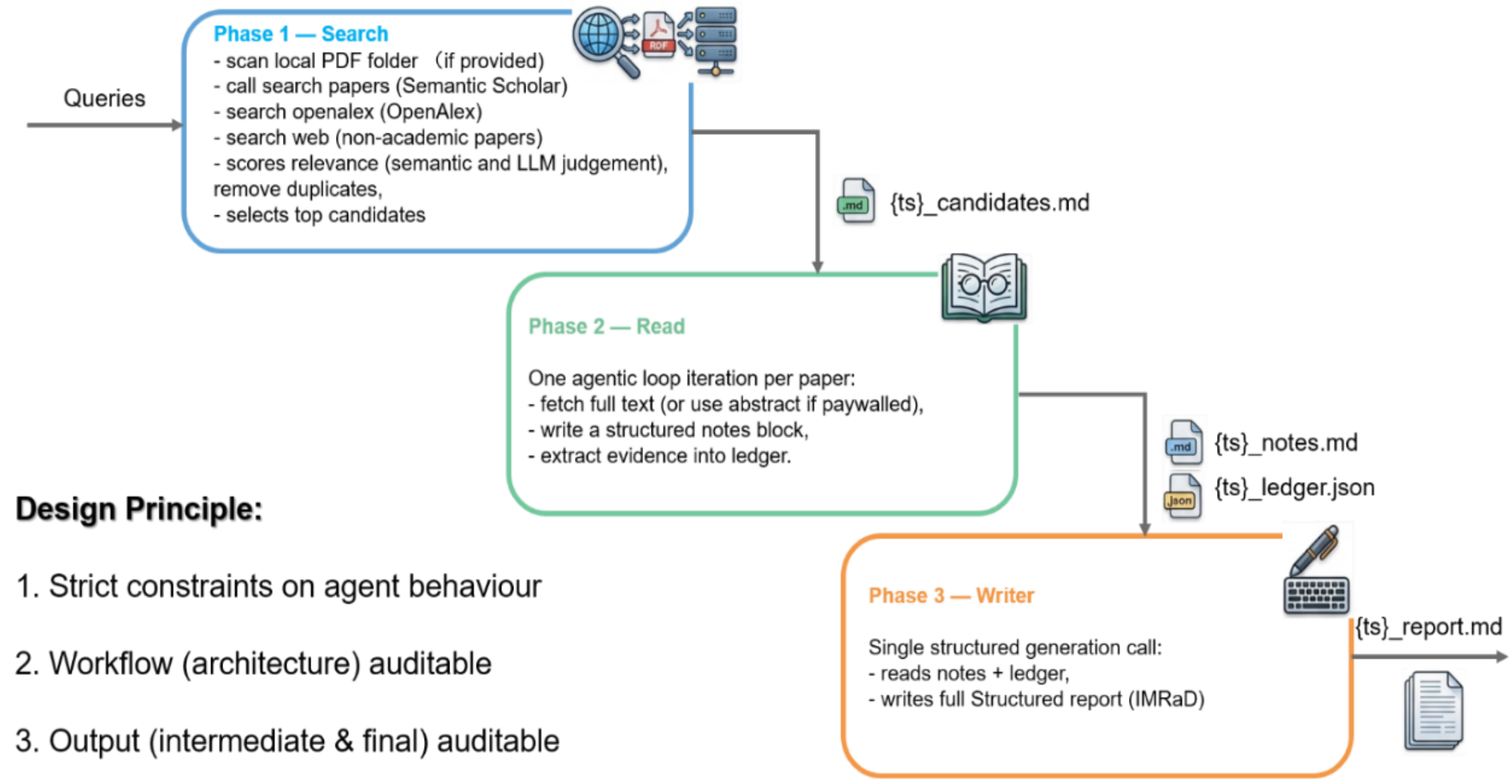

Figure A1. The research workflow embedded in the contextualization agent and expert agents

**Phase 1: Search**. Following the entrance of a research question/topic, the search agent starts to generate queries from different angles to identify relevant information. The sources are ranked by semantic similarity to the research question. The top N candidates are selected and further scrutinised to exclude those that are semantically similar but contextually mismatched (e.g., wrong geographical area or time scope). The final output of the search agent is a Markdown file with the metadata about the most relevant candidates.

**Phase 2: Read.** A reading agent is responsible for fetching and reading the full text of the candidates to extract useful evidence and claims. The reading agent prioritises sources including (1) designated local directory, (2) full-text links provided by Semantic Scholar, (3) paper DOIs if unpaywalled, and (4) abstracts and titles in the metadata if none of the other approaches works. The outcome of the reading agent includes a “ledger” in the JSON format, containing four categories of information: information sources, extracted evidence, claims, and assumptions. Each specific item under these categories is labelled uniquely. The information collected and sorted is stored and made available for further scrutiny and analysis beyond WEGA.

**Phase 3: Write.** The writer agent takes the outputs from the reading agent and generates a full report with the rigidity of an academic article, meaning that the entire process of agentic reasoning becomes auditable: identification of the gaps given the state of the art, the logic flow regarding the key causal relationships in the system, evidence, and the logic flow the writer agent uses to form conclusions. The final report can be audited via a web-based UI with Title, Abstract, Introduction, Methodology, Findings, Discussion, Assumptions, Conflicts, Conclusion, and References sections supported by the evidence cited properly. It is noteworthy that Phase 3 can be skipped to save time and computing resources. Because the outputs from Phase 2 provide sufficient “building blocks” to construct context and make evidence-based claims. The output of “Write” provides a more human-friendly format to scrutinise and audit the three phases, as it converts the “ledger” in JSON format to a paper-like narrative.

## Appendix B

Table B1. Stakeholder decision sets generated by WEGA

**<u>Decision set 1</u>**

**NNRP_2026_RATIFICATION**

The Dutch Cabinet ratified the National Nitrogen Transition & Compliance Framework (2026–2030) to be submitted as the legally binding National Nitrogen Reduction Plan (NNRP) to the European Commission.

**REJECTION_OF_MORATORIA**

The government formally rejected all requests for moratoria on court-enforced reduction obligations, permit freezes, or compliance timelines, affirming that ecological thresholds do not pause for political calibration.

**SUPPLY_CHAIN_EPR_LEVY**

Establishment of a Supply-Chain Extended Producer Responsibility (EPR) Levy on meat/dairy processors and synthetic fertilizer importers to fund the nitrogen transition, shifting financial burden away from sole reliance on primary producers.

**TRANSITION_FUNDING_ALLOCATION**

Mobilization of €2.4 billion in ring-fenced funding (via CAP Pillar 2 and national budget) providing grants and low-interest loans covering 70–80% of capital costs for precision feeding, low-emission housing, and manure separation, contingent on verified surplus reduction.

**OPEN_DATA_MANDATE**

Mandatory publication of all raw deposition modelling code, calibration datasets, and critical load assumptions through an independent scientific repository to ensure methodological transparency while maintaining compliance deadlines.

**DIFFERENTIATED_ZONE_TARGETS**

Implementation of zone-specific reduction targets calibrated to critical load thresholds, soil buffering capacity, and hydrological lag times, replacing blanket herd culls with compliance measured against annual verified surplus reductions.

**MANDATORY_SECTOR_TRANSPARENCY**

Introduction of mandatory, standardized reporting on manure routing, feed nitrogen composition, and on-farm leaching metrics integrated into national monitoring systems, made a condition for permit retention and subsidy access.

**<u>Decision set 2</u>**

**SNTA-01**

Enactment of the Structured Nitrogen Transition Accord (SNTA), establishing a conditional framework replacing punitive paralysis with legally compliant implementation.

**SNTA-02**

Implementation of a 30-day monitoring restoration period requiring cessation of equipment interference and reporting falsification, verified by independent third-party audit, as a precondition for a subsequent 90-day conditional transition window.

**SNTA-03**

Establishment of the Joint Nitrogen Attribution & Mitigation Review (JNAMR), a statutory co-chaired consortium to stress-test deposition models and refine source attribution, with strict guardrails preventing baseline recalibration for delay or directive overrides.

**SNTA-04**

Suspension of the Supply-Chain EPR Levy and its replacement with the Dutch-EU Agricultural Transition Resilience Fund, providing conditional grants for technology upgrades and debt restructuring tied to verified environmental milestones.

**SNTA-05**

Pause on new administrative herd-reduction orders and zone-specific moratoria during the 90-day negotiation window, while maintaining existing legal reduction trajectories and shifting compliance tracking to quarterly outcome-based pathways.

**Decision set 3**

**SNTA-2036-DEC-01**

Enactment of a 'Trajectory-Conditional Structural Guarantee' under the SNTA Addendum. Mandatory herd reductions, state buyouts, and culling triggers are suspended provided the Netherlands maintains its verified biannual deposition trajectory toward Natura 2000 critical loads. Compliance metrics shift from livestock headcounts to emission-outcome metrics. A proportional, co-managed adjustment protocol will activate only if habitat-level deposition plateaus across two consecutive MRV cycles.

**SNTA-2036-DEC-02**

Accelerated deployment of the next Resilience Fund tranche synchronized with CAP eco-schemes (up to 70% co-financing) for precision feeding, closed-loop manure processing, and ammonia capture. Concurrently, 30% of this capital is ringfenced specifically for legacy groundwater remediation (riparian buffers, controlled drainage, constructed wetlands, soil carbon restoration). Disbursements are milestone-tied but legally firewalled from retroactive clawbacks for non-systemic reporting deviations.

**SNTA-2036-DEC-03**

Mandate for JNAMR model re-calibration and open-source architecture. The RIVM, in coordination with the JRC and EEA, will publish full JNAMR source code, atmospheric dispersion assumptions, and weighting parameters for open peer review. Legacy groundwater nitrogen will be administratively decoupled from current operational accounting pending peer-reviewed hydrological validation. Industrial, transport, and urban emissions will be integrated into a unified National Nitrogen Balance Dashboard with identical real-time MRV standards.

**SNTA-2036-DEC-04**

Institution of legally binding biannual ecological MRV checkpoints measuring actual habitat-level critical load attainment at Natura 2000 sites. The independent monitoring data from the NGO/Scientist Coalition (satellite $NH_3$ tracking, biodiversity metrics) is formally integrated into the national EEA-MRV architecture as a statutory oversight mechanism.

**SNTA-2036-DEC-05**

Introduction of legislation requiring Independent Rural Impact Assessments and a two-thirds parliamentary majority for any ecological set-aside or zoning overlay intersecting with active agricultural land. Existing agricultural permits are shielded from automatic nullification for SNTA-compliant farms. Livestock farming is formally recognized in the SNTA addendum as a protected socio-economic function with statutory cultural continuity provisions.

**Decision set 4**

**SNTA-01**

Enactment of the Structured Nitrogen Transition Accord (SNTA), establishing a conditional framework replacing punitive paralysis with legally compliant implementation.

**SNTA-02**

Implementation of a 30-day monitoring restoration period requiring cessation of equipment interference and reporting falsification, verified by independent third-party audit, as a precondition for a subsequent 90-day conditional transition window.

**SNTA-03**

Establishment of the Joint Nitrogen Attribution & Mitigation Review (JNAMR), a statutory co-chaired consortium to stress-test deposition models and refine source attribution, with strict guardrails preventing baseline recalibration for delay or directive overrides.

**SNTA-04**

Suspension of the Supply-Chain EPR Levy and its replacement with the Dutch-EU Agricultural Transition Resilience Fund, providing conditional grants for technology upgrades and debt restructuring tied to verified environmental milestones.

**SNTA-05**

Pause on new administrative herd-reduction orders and zone-specific moratoria during the 90-day negotiation window, while maintaining existing legal reduction trajectories and shifting compliance tracking to quarterly outcome-based pathways.